\documentclass[prl,preprint,twocolumn,10pt,tightenlines]{revtex4}
\usepackage{graphicx}
\usepackage{bm}

\begin{document}

\preprint{NT@UW-03-015}
\title{Chiral solitons in nuclei: Saturation, EMC effect and Drell-Yan experiments}
\author{Jason R. Smith and Gerald A. Miller}
\affiliation{Department of Physics\\University of
Washington\\Seattle, WA 98195-1560}
\begin{abstract}
The Chiral Quark-Soliton model of the nucleon contains a mechanism
for an attractive interaction between nucleons. This, along with
the exchange of vector mesons between nucleons, is used to compute
the saturation properties of infinite nuclear matter. This
provides a new way to asses the effects of the nuclear medium on a
nucleon that includes valence and sea quarks. We show that the
model simultaneously describes the nuclear EMC effect and the
related Drell-Yan experiments.
\end{abstract}

\maketitle

One frontier of strong interaction physics lies in the
intermediate range of length scales available to present
experiments where neither the fundamental theory, Quantum
Chromodynamics (QCD), nor its low energy effective theory, Chiral
Perturbation Theory, have useful perturbative expansions. Neither
fundamental quarks nor point-like hadrons provide a complete
description, so including the non-perturbative information that
hadrons are bound states of valence quarks in a polarized vacuum
is necessary. One way to probe these intermediate length scales
and this non-perturbative physics is to examine the short distance
structure of a large object. The prime example is the
European Muon Collaboration (EMC) effect \cite{Aubert:1983xm}
where the short distance ($\sim 5$ GeV, or $\sim 10^{-2}\text{
fm}$) structure of nuclei differs from that of
a collection of free nucleons. This measurement showed that bound
nucleons are different than free ones, and implied that the medium
modifications could be significant for any nuclear observable
\cite{Geesaman:1995yd}. Indeed, a recent paper \cite{str} obtains
evidence for a medium modification of the elastic proton form
factor.

Our central concern is the depletion of the nuclear structure
function $F_{2}^{A}(x)$ in the valence quark regime $0.3\lesssim x
\lesssim 0.8$. While the general interpretation is that a valence
quark in a bound nucleon has less momentum than in a free one,
corresponding to some increased length scale, the specific
mechanism for this has eluded a universally accepted explanation
for 20 years
\cite{Geesaman:1995yd,Piller:1999wx,Arneodo:1992wf,Sargsian:2002wc}.
A popular explanation is the so-called `binding' effect which
originates from a possible mechanism in which mesons binding the
nucleus carry momentum. An important consequence is that the
mesonic presence would enhance the anti-quark content of the
nucleus \cite{bickerstaff:1984,ericson:1984}. Such an effect has
not been seen in Drell-Yan experiments \cite{Alde:im} in which a
quark in a proton beam annihilates with an antiquark in a nuclear
target producing a muon pair. Furthermore, relativistic
treatments, including the light-cone approach needed to obtain the
nucleon structure function, of the binding effect with
structureless hadrons fail
\cite{Miller:2001tg,Smith:2002ci,Birse:hu,Frankfurt:1985ui},
suggesting that modifications of the internal quark structure of
the nucleon are required to explain the deep inelastic scattering
data.

Any description of the EMC effect must be consistent with the
constraints set by both deep inelastic scattering and Drell-Yan
data. Thus a successful model must include antiquarks as well as
quarks, and show how the medium modifies both the valence and sea
quark distributions. Our purpose is to provide a
mechanism for that modification within the Chiral Quark-Soliton
(CQS) model
\cite{Kahana:dx,Birse:1983gm,Diakonov:2000pa,Christov:1995vm}.
This phenomenological model has many desirable qualities: the
ability to describe a wide class of hadron observables with
surprising accuracy, the
inclusion of antiquarks, positivity of Generalized Parton
Distributions, and a basis in QCD \cite{Diakonov:2000pa}. The
model also predicted \cite{Diakonov:1997mm} the recently
discovered $\theta^{+}$ exotic baryon resonance
\cite{Barmin:2003vv,Nakano:2003qx}. Here we show how the model
describes nuclear saturation properties, reproduces the EMC
effect, and satisfies the bounds on nuclear antiquark enhancement
provided by Drell-Yan experiments.

The CQS model Lagrangian with (anti)quark fields
$\overline{\psi},\psi$, and profile function $\Theta(r)$ is
\begin{equation}
\mathcal{L} =  \overline{\psi} ( i \partial \!\!\!\!\!\:/\, - M
e^{  i \gamma_{5}\bm{n}\cdot\bm{\tau} \Theta(r) } ) \psi,
\label{eq:lagrangian}
\end{equation}
where $\Theta(r\rightarrow\infty) = 0$ and $\Theta(0) = -\pi$
to produce a soliton with unit winding number. The quark
spectrum consists of a single bound state and a filled negative
energy Dirac continuum; the vacuum is the filled negative
continuum with $\Theta = 0$. The wave functions in this spectrum
provide the input for the quark and antiquark distributions used
to calculate the nucleon structure function.

We work to leading order in the number of colors ($N_{C}=3$), with
$N_{f}=2$, and in the chiral limit. While the former characterizes
the primary source of theoretical error, one could systematically
expand in $N_{C}$ to calculate corrections. We take the
constituent quark mass to be $M=420\text{ MeV}$, which reproduces,
for example, the $N$-$\Delta$ mass splitting at higher order in
the $N_{C}$ expansion, and other observables
\cite{Christov:1995vm}.

The theory contains divergences that must be regulated. We use a
single Pauli-Villars subtraction as in Ref.~\cite{Diakonov:1997vc}
because we follow that work to calculate the quark distribution
functions. The Pauli-Villars mass is determined by reproducing the
measured value of the pion decay constant, $f_{\pi} = 93 \text{
MeV}$, with the relevant divergent loop integral regularized using
$M_{PV}\simeq 580 \text{ MeV}$. This regularization also preserves
the completeness of the quark states \cite{Diakonov:1997vc}.

The nucleon mass is given by a sum of the energy of a single
valence level ($E^{v}$), and the regulated energy of the soliton
($E_{\Theta}$, equal to the energy in the negative Dirac continuum
with the energy in the vacuum subtracted)
\begin{eqnarray}
M_{N} = N_{C} E^{v}+
E_{\Theta}(M)-\frac{M^{2}}{M_{PV}^{2}}E_{\Theta}(M_{PV}).\label{eq:mn}
\end{eqnarray}
The field equation for the profile function is
\begin{equation}
\Theta(r) = \arctan
\frac{\rho_{ps}^{q}(r)}{\rho_{s}^{q}(r)},\label{eq:thetafe}
\end{equation}
where $\rho_{s}^{q} \text{ and } \rho_{ps}^{q}$ are the quark
scalar and pseudoscalar densities, respectively.

The dependence of nucleon properties on the nuclear medium is
incorporated in the model by simply letting the quark scalar
density in the field equation (\ref{eq:thetafe}) contain a
(constant) contribution arising from other nucleons present in
symmetric nuclear matter. This models a scalar interaction via the
exchange of multiple pairs of pions between nucleons. We take the
scalar density to consist of three terms: 1) the constant
condensate value $\langle \overline{\psi}\psi\rangle_{0}$ (in the
vacuum or at large distances from a free nucleon), 2) the valence
contribution $\rho_{s}^{v}$ and 3) the contribution from the
medium which takes the form of the convolution of the nucleon
$\rho_{s}^{N}$ and valence quark scalar densities as in the
Quark-Meson Coupling (QMC) model
\cite{Saito:ki,Saito:yw,Blunden:1996kc}
\begin{equation}
\rho_{s}^{q}(r)  \simeq  \langle \overline{\psi}\psi\rangle_{0} +
\rho_{s}^{v}(r) + \int d^{3}r'\rho_{s}^{N}(r-r')\rho_{s}^{v}(r').
\label{eq:rhos}
\end{equation}
We take the pseudoscalar density to have only the valence term
$\rho_{ps}^{q} \simeq \rho_{ps}^{v}$; the two other contributions
analogous to the first and third terms of Eq.~(\ref{eq:rhos})
vanish due to symmetries of the QCD vacuum and nuclear matter.
These approximations to the densities neglect the precise form of
the negative continuum wave functions in Eq.~(\ref{eq:thetafe}).
The resulting free nucleon profile function has no discernible
difference from a fully self-consistent treatment, demonstrating
the excellence of this approximation.

We take the vacuum value of the chiral condensate in
Eq.~(\ref{eq:rhos}) to be free parameter, but in the single
Pauli-Villars regularization, the scalar and pseudoscalar
densities contain a divergence that cancels in the ratio
Eq.~(\ref{eq:thetafe}) \cite{Kubota:1999hx}. A finite value of the
ratio is obtained by normalizing the densities so that the
cancellation occurs (by dividing the other terms in the numerator
and denominator by the same divergent quantity). This yields a
free nucleon mass that is independent of $\langle
\overline{\psi}\psi\rangle_{0}$ (as is necessary because it is
divergent in single Pauli-Villars regularization), and a medium
contribution that enters only through the ratio $\rho_{s}^{N}
/\langle \overline{\psi}\psi\rangle_{0}$. While the vacuum value
of the condensate does not vary by definition, the effective
condensate $\langle \overline{\psi}\psi\rangle_{0}+\rho_{s}^{N}S(k_{F})$,
where $S(k_{F})$ is integral of $\rho_{s}^{v}$ (see Eq.~(\ref{eq:rhos})),
falls $\sim 30 \%$ at nuclear density. This is consistent with the model
independent result \cite{Cohen:1991nk} that predicts a value 25-50\%
below the vacuum value.

The nucleon scalar density is determined by solving the nuclear
self-consistency equation
\begin{equation}
\rho_{s}^{N} = 4 \int^{k_{F}} \frac{d^{3}k}{(2\pi)^{3}}
\frac{M_{N}(\rho_{s}^{N})}{\sqrt{k^{2}+M_{N}(\rho_{s}^{N})^{2}}}.\label{eq:nsc}
\end{equation}
The dependence of the nucleon mass, and any other properties
calculable in the model, on the Fermi momentum $k_{F}$ enters
through Eq.~(\ref{eq:nsc}). Thus there are two coupled
self-consistency equations: one for the profile,
Eq.~(\ref{eq:thetafe}), and one for the density,
Eq.~(\ref{eq:nsc}). These are iterated until the change in the
nucleon mass Eq.~(\ref{eq:mn}) is as small as desired for each
value of the Fermi momentum. We use the Kahana-Ripka (KR) basis
\cite{Kahana:be}, with momentum cutoff and box size extrapolated
to infinity, to evaluate the energy eigenvalues and wave functions
used as input for the densities, nucleon mass, and quark
distributions.

A phenomenological vector meson (mass $m_{v}=770\text{ MeV}$)
exchanged between nucleons (but not quarks in the same nucleon), is
introduced \cite{Walecka:qa}
to obtain the necessary short distance repulsion which stabilizes
the nucleus. The resulting energy per nucleon is
\begin{equation}
\frac{E}{A} = \frac{4}{\rho_{B}(k_{F})} \int^{k_{F}}
\frac{d^{3}k}{(2\pi)^{3}} \sqrt{k^{2} + M_{N}(k_{F})^{2}}
+\frac{1}{2}\frac{g_{v}^{2}}{m_{v}^{2}}\rho_{B}(k_{F})
\label{eq:epn}.
\end{equation}

\begin{figure}
\centering
\includegraphics[scale=0.5]{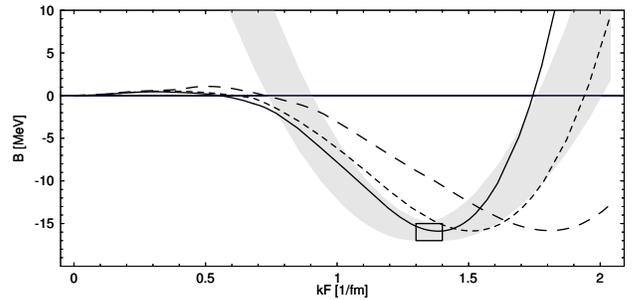}
\caption{Binding energy per nucleon as a function of Fermi
momentum fit to $B=-15.75\text{ MeV}$ at the minimum for $\langle
\overline{\psi}\psi \rangle_{0} = -(225 \text{ MeV})^{3}$ (long
dashed), $-(210 \text{ MeV})^{3}$ (short dashed) and $-(200 \text{
MeV})^{3}$ (solid). The box and shaded region are the experimental
uncertainty \cite{Blaizot:tw} in the binding energy, density and
compressibility of nuclear matter.} \label{fig:bepn2}
\end{figure}

We now present the results. The mass of a free nucleon is computed
to be $M_{N}(k_{F}=0)=1209\text{ MeV}$. The $\sim 30\%$ difference
is as expected in the model at leading order in $N_{C}$.  We
evaluate the nucleon mass Eq.~(\ref{eq:mn}) and energy per nucleon
Eq.~(\ref{eq:epn}) as a function of $k_F$ for three values of the
condensate. We plot $B = E/A - M_{N}(0)$ in Fig.~\ref{fig:bepn2}
where we choose the vector meson coupling to fit $B = -15.75
\text{ MeV}$ at the minimum.

We use the value $-\langle \overline{\psi}\psi \rangle_{0}
^{1/3}=200\text{ MeV}$, and vector coupling $g_{v}^{2}/4\pi = 10.55$,
which gives a Fermi momentum of $k_{F} =
1.38\text{ fm}^{-1}$ in nuclear matter consistent with the known
value $k_{F} = 1.35 \pm 0.05 \text{ fm}^{-1}$ \cite{Blaizot:tw}.
The compressibility is $K = 348.5\text{ MeV}$ which is above the experimental value $K =
210 \pm 30 \text{ MeV}$, but well below the Walecka model value of
$560 \text{ MeV}$.

The isoscalar unpolarized distribution $q(x)= u(x)+d(x)$ is the
leading order term in $N_{C}$, with the isovector unpolarized
quark distribution $u(x)-d(x)$ smaller by a factor $ \sim 1/N_c$
and set to zero. The distributions are calculated using the KR
basis at $k_{F}=0$ and $k_{F}=1.38\text{ fm}^{-1}$ almost exactly
as in Ref.~\cite{Diakonov:1997vc} where the quark distribution is
given by the matrix element
\begin{equation}
q(x) = N_{C} M_{N} \sum_{n} \langle \psi_{n} |
(1+\gamma^{0}\gamma^{3})\delta(E_{n}+p^{3}-x M_{N}) | \psi_{n}
\rangle,\label{eq:me}
\end{equation}
with the regulated sum taken over occupied states. The eigenvalues
$E_{n}$ are determined from diagonalizing the Hamiltonian, derived
from the Lagrangian (\ref{eq:lagrangian}), in the KR basis.
The vector meson exchange is not explicit in Eq.~(\ref{eq:me}) because
the initial, intermediate and final states of the struck and spectator
quarks experience the same vector potential, as demanded by consistency.
Thus we include the interaction of the debris of the struck nucleon with
the residual nucleus \cite{Saito:yw}.
The antiquark distribution is given
by $\bar{q}(x) = -q(-x)$ where the sum is over unoccupied states.
We use the exact sea wave functions, and not the approximation
used in Eq.~(\ref{eq:rhos}). The use of a finite basis causes the
distributions to be discontinuous. These distributions are smooth
functions of $x$ in the limit of infinite momentum cutoff and box
size, but numerical calculations are made at finite values and
leave some residual roughness. This is overcome in
Ref.~\cite{Diakonov:1997vc} by introducing a smoothing function.
We deviate from their procedure, and do not smooth the results;
instead we find the subsequent one-loop perturbative QCD evolution
\cite{Hagiwara:fs} to be sufficient.

\begin{figure}
\centering
\includegraphics[scale=0.5]{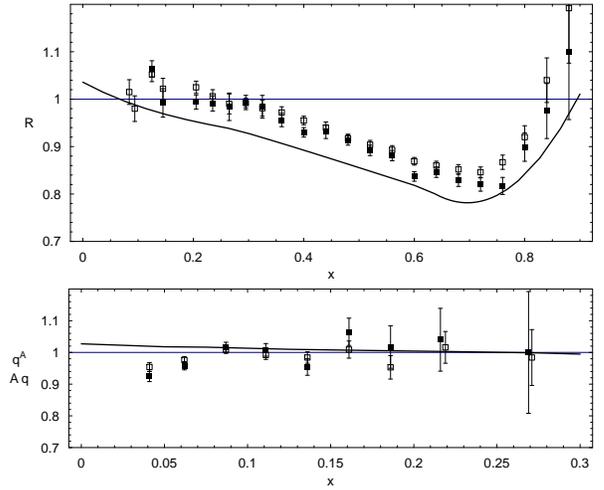}
\caption{The EMC (top) and the Drell-Yan (bottom) ratios at scales
$Q = 5 \text{ GeV}$ and $10 \text{ GeV}$, respectively, for nuclear
matter. The data are for Iron (empty boxes) and Gold (filled
boxes) from SLAC-E139 (top) \cite{Gomez:1993ri} for Iron (empty
boxes) and Tungsten (filled boxes) from FNAL-E772 (bottom)
\cite{Alde:im}.}\label{fig:emcdy}
\end{figure}

These distributions are used as input at a scale of $Q =
M_{PV}\simeq 580 \text{ MeV}$ for evolution to $Q = 5 \text{ GeV}$
in the case of the quark singlet distribution
$q^{S}(x)=q(x)+\bar{q}(x)\propto F_{2}^{N}(x)/x$ at leading order
in $N_{C}$. The EMC ratio function is defined to be
\begin{eqnarray}
R(x,Q^{2}) & = & \frac{F_{2}^{A}(x,Q^{2},k_{F})}{A
F_{2}^{N}(x,Q^{2},k_{F}=0)},\label{eq:ratio}\\
F_{2}^{A}(x,Q^{2},k_{F}) & = & \int_{x}^{A} dy f(y)
F_{2}^{N}(x/y,Q^{2},k_{F}).\nonumber
\end{eqnarray}
The nucleon momentum distribution, following from a light-cone
approach, for any mean field theory of nuclear matter
\cite{Miller:2001tg} is
\begin{equation}
f(y) = \frac{3}{4\Delta_{F}^{3}} \theta(1+\Delta_{F}-y)
\theta(y-1+\Delta_{F}) \left[
\Delta_{F}^{2}-(1-y)^2\right],\label{eq:pmd}
\end{equation}
where $\Delta_{F} = k_{F}/\overline{M}_{N}$ and
$\overline{M}_{N}=M_{N}(0)-15.75 \text{ MeV}$. The antiquark
distribution $\bar{q}(x)$ is evolved to $Q = 10 \text{ GeV}$ for
use in the Drell-Yan ratio $\bar{q}^{A}/A \bar{q}$, analogous to
Eq.~(\ref{eq:ratio}). The EMC and Drell-Yan ratios are plotted in
Fig.~\ref{fig:emcdy}. While the data shown in Fig.~\ref{fig:emcdy}
are for large, but finite, nuclei, our calculation reproduces the
trend of both sets of data. It falls slightly below the SLAC-E139
data \cite{Gomez:1993ri} due to the higher density of nuclear
matter.

\begin{figure}
\centering
\includegraphics[scale=0.5]{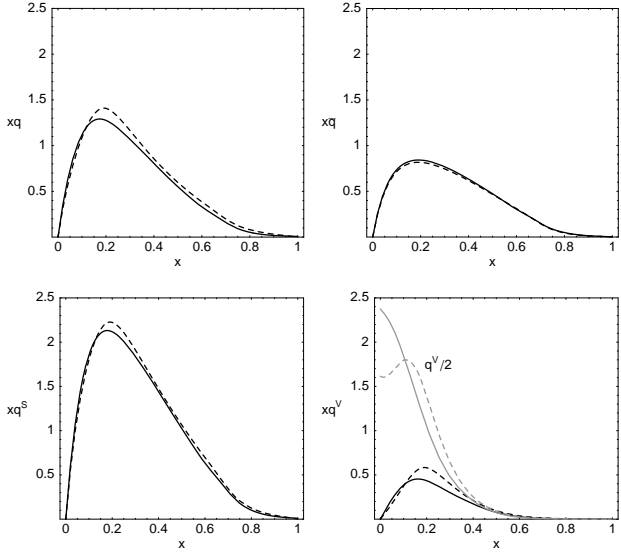}
\caption{Clockwise from the top left the distributions
$xq(x),x\bar{q}(x),xq^{V}(x),xq^{S}(x)$ in a free (dashed) and
bound (solid) nucleon at a scale $Q = 5 \text{ GeV}$. The valence
distribution $q^{V}(x)/2$ is also shown in gray in the lower right
graph.} \label{fig:dists}
\end{figure}

In Fig.~\ref{fig:dists} we show the quark, antiquark, singlet and
valence ($q^{v} = q-\bar{q}$) quark distributions weighted by $x$
for a free and bound nucleon at a scale $Q=5\text{ GeV}$.
There is a large depletion in the bound nucleon valence
distribution in Fig.~\ref{fig:dists}, that, if used to calculate
the EMC ratio (\ref{eq:ratio}), produces too large an effect. This
large effect is comparable to that of the QMC model impulse
approximation calculation or the Guichon model \cite{Saito:yw}
which only include valence quarks. This valence effect is
mitigated by a small enhancement in $x\bar{q}$, consistent with
the Drell-Yan data, so that the singlet distribution has only a
moderate depletion consistent with the EMC effect.

A simple picture in terms of the uncertainty principle is
available. The influence of the nuclear medium on the nucleon
causes the root mean square radius of the baryon density to
increase by 2.4\%. This corresponds to a decreased momentum, and a
depletion of the bound structure function relative to the free
one. This swelling is consistent with a $<6 \%$ increase as
constrained by quasi-elastic inclusive electron-nucleus scattering
data \cite{Mckeown:kn}, and the recent polarization transfer
measurement \cite{str}.

We ignore the effects of shadowing, which occur when the virtual
photon striking the nucleus fluctuates into a quark-antiquark pair
over a distance $\sim 1/2 M_{N} x$ exceeding the inter-nucleon
separation. This causes a depletion in the structure function for
$x\lesssim 0.1$ and is relatively well understood
\cite{Geesaman:1995yd,Piller:1999wx,Arneodo:1992wf,Sargsian:2002wc}
and so we do not reiterate those results. Additionally, we ignore
contributions from quantum pion structure functions, which in this
model propagate through constituent quark loops, and would modify
the behavior at small $x$. These loops are suppressed by
${\mathcal O}(1/N_{C})$, and are not treated at leading order.

The present model provides a intuitive, qualitative treatment that
maintains consistency with all of the free nucleon properties
calculated by others \cite{Diakonov:2000pa,Christov:1995vm}. It
gives a reasonable description of nuclear saturation properties,
reproduces the EMC effect, and satisfies the constraints on the
nuclear sea obtained from Drell-Yan experiments with only two free
parameters: $\langle \overline{\psi} \psi \rangle_{0}$ and
$g_{v}$.

The central mechanism to explain the EMC effect is that the
nuclear medium provides an attractive scalar interaction that
modifies the nucleon wave function. This is also the dominant
mechanism in the QMC model approach to the EMC effect
\cite{Saito:yw} and also similar to the quark delocalization
approach \cite{Benesh:2003fk}. The improvements given here are the
explicit computation of the effects of the medium on the antiquark
distributions so that consistency with the Drell-Yan data could be
verified, and the reduction of the number of input parameters and
model assumptions. Our extension of the Chiral Quark-Soliton model
to nuclear matter provides a new, consistent way to calculate
possible medium modifications of a variety of observables that
could be measured in experiments.

We would like to thank the USDOE for partial support of this work,
and S.~D.~Ellis for useful comments.

\end{document}